\documentclass[english,aps,prb,showpacs,twocolumn,superscriptaddress]{revtex4}
\usepackage{amsmath}
\usepackage{amsfonts}
\usepackage{amssymb}
\usepackage{graphicx}

\usepackage[latin9]{inputenc}
\usepackage{babel}
\usepackage{xspace}
\usepackage{color}
\usepackage{colordvi}

\begin{document}
\title{Optimal control of circuit quantum electrodynamics in one and two dimensions}
\author{R. Fisher}
\email{fisher@tum.de}
\affiliation{Department of Chemistry, Technische Universit\"at M\"unchen, Lichtenbergstrasse 4, D-85747 Garching, Germany}
\author{F. Helmer}
\affiliation{Department of Physics, Center for NanoScience, and Arnold Sommerfeld Center for Theoretical Physics, Ludwig Maximilians Universit\"at M\"unchen, Theresienstrasse 37, D-80333 Munich, Germany}
\author{S. J. Glaser} 
\affiliation{Department of Chemistry, Technische Universit\"at M\"unchen, Lichtenbergstrasse 4, D-85747 Garching, Germany}
\author{F. Marquardt}
\affiliation{Department of Physics, Center for NanoScience, and Arnold Sommerfeld Center for Theoretical Physics, Ludwig Maximilians Universit\"at M\"unchen, Theresienstrasse 37, D-80333 Munich, Germany}
\author{T. Schulte-Herbr\"uggen}
\affiliation{Department of Chemistry, Technische Universit\"at M\"unchen, Lichtenbergstrasse 4, D-85747 Garching, Germany}
\begin{abstract}
Optimal control can be used to significantly improve multi-qubit gates in quantum information processing hardware architectures based on superconducting circuit quantum electrodynamics. We apply this approach
not only to dispersive gates of two qubits inside a cavity, but, more generally, to architectures based on two-dimensional arrays of cavities and qubits. For high-fidelity gate operations, {\em simultaneous} evolutions of controls and couplings in the two coupling dimensions of cavity grids are shown to be significantly faster than conventional {\em sequential} implementations. Even under experimentally realistic conditions speedups by a factor of three can be gained. The methods immediately scale to large grids and indirect gates between arbitrary pairs of qubits on the grid. They are anticipated to be paradigmatic for 2D arrays and lattices of controllable qubits.
\end{abstract}
\date{\today}
\maketitle
\newcommand{\ket}[1]{\left|#1\right\rangle }
\newcommand{\bra}[1]{\left\langle #1\right|}
\newcommand{\s}{\hat{\sigma}}
\newcommand{\tr}{\operatorname{tr}\xspace}
\renewcommand{\Re}{\operatorname{Re}}
\newcommand{\unity}{\ensuremath{{\rm 1 \negthickspace l}{}}}
\newcommand{\Adr}{\operatorname{Ad}}
\newcommand{\adr}{\operatorname{ad}}
\section{Introduction}
Progress towards the goal of scalable quantum information processing is currently concentrated in physical systems that live at the intersection between quantum optics and solid state physics. One of the most promising contenders are superconducting circuits that couple qubits and microwave resonators, a new field now known as circuit quantum electrodynamics (QED). In the development of this field, early suggestions to implement the Jaynes-Cummings model in the solid state \cite{Marquardt2001, BuissonHekking2001, YouNori2003} were followed by a seminal proposal \cite{Blais2004} to employ on-chip microwave resonators and couple them to artifical atoms in the form of superconducting qubits. Experimental realizations soon followed\cite{Wallraff2004}, creating a solid-state analogue of conventional optical cavity QED \cite{CavQEDReview}. The tight confinement of the field mode and the large electric dipole moment of the {}`atom' yield extraordinary coupling strengths, which has led to a variety of experimental achievements, including: The Jaynes-Cummings model in the strong-coupling regime \cite{Wallraff2004, Mooij2004, Semba2006}, Rabi and Ramsey oscillations and dispersive qubit readout \cite{Schuster2006, WallraffPRL2005}, generation of single photons \cite{MicrowaveSource2007} and Fock states \cite{hofheinz2008gfs, wang2008mdf}, cavity-mediated coupling of two qubits \cite{majer2007csq, Simmonds2007}, setups with three qubits \cite{Wallraff3Qubits2009}, Berry's phase \cite{leek2007obs}, and the measurement of the photon number distribution \cite{YalePhotonNumberSplitting2007}.

The recent experimental progress in creating microwave circuits with multiple qubits coupled to resonators establishes the need for efficient, high-fidelity multi-qubit quantum gates and motivates the search for
advanced architectures for quantum information processing on the chip. The most elementary situation to consider is two qubits inside a cavity (transmission line resonator). The mere presence of the cavity induces a flip-flop ($XY$) type interaction between the qubits, which may and has been used for entangling gate operations\cite{majer2007csq}. The $XY$ interaction directly produces an i{\sc swap} two-qubit gate. Other gates (like the {\sc cnot}) have to be synthesized. While it is well-known how to do so using a sequence of i{\sc swap} and single-qubit gates \cite{SchuchSiewert,CavityGridEPL}, there is a lot of room for improvement in constructing faster gates even for this basic situation. One way to go is to use resonant two-qubit gates \cite{Yale2Qubit2009,Haack2009}. The other approach is to keep the robust dispersive $XY$ interaction and to explore better pulse sequences using the tools of optimal control theory. This is the approach we will follow in this paper.

\begin{figure}[!htbp]
\includegraphics[width=0.97\columnwidth]{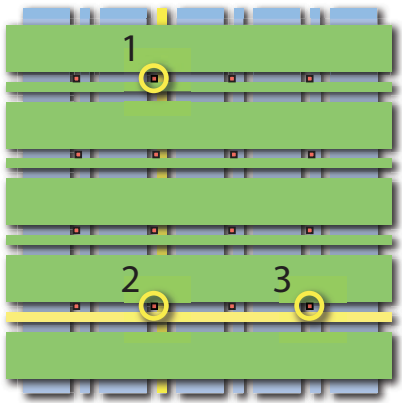}
\caption{(Color online) The superconducting cavity grid\cite{CavityGridEPL}, with two layers of vertical (bottom) and horizontal (top layer) transmission line resonators, coupled to qubits (small red squares). Two-qubit gates between qubits 1 and 3 are mediated indirectly via qubit 2, employing the dispersive interaction inside the two highlighted resonators.}
\label{cavGrid}
\end{figure}

When going beyond two qubits and connecting many qubits into a quantum processor on a chip, it is crucial to abandon linear arrays and to extend the setup into the second dimension. While there have been a number of schemes for doing so in nearest-neighbor coupled 2D arrays, the presence of global coupling between qubits via resonators adds a new feature that has to be explored. Perhaps the most straightforward route, recently introduced by some of us \cite{CavityGridEPL}, is to create a two-dimensional {}`cavity grid' of transmission line resonators arranged in columns and rows (Fig.~\ref{cavGrid}).
The qubits are located at the intersections, such that each qubit feels the microwave field of two cavities. In this way, qubits can be addressed and coupled. An interesting feature specific to such a global coupling architecture is the fact that two qubits placed anywhere on the grid can be coupled via a third qubit sitting at the intersection of two resonators, with an overhead that does not grow with system size. Again, while a sequential protocol for this {}`coupling around the corner' is known \cite{CavityGridEPL}, we may ask for possible speedups acquired by more sophisticated pulse schemes.

Principles of optimal quantum control \cite{Sam90} are currently establishing themselves as indispensible tools to steer quantum systems in a robust, time-optimal or loss-avoiding way{\cite{DowMil03}}. Earlier examples of numerical optimal control can be found in Ref.~\onlinecite{Rabitz87}. Based on the gradient ascent approach {\sc grape}\cite{GRAPE}, our toolbox of optimal control techniques now allows the synthesis of quantum gates in a time-optimal\cite{PRA05} or relaxation-optimized way, where the open system may come in a Markovian \cite{PRA_decoh} or a non-Markovian setting \cite{PRL_decoh2}.

In particular in superconducting qubits (recently reviewed in Ref.~\onlinecite{CW08}), the issue of improving qubit gates by optimal control techniques is currently attracting increased attention (see Refs. \onlinecite{Spoerl2007, Rebentrost2009, Safaei2009, Roloff2009, Motzoi2009, Jirari2009} for some of the approaches). Here, we take these ideas further by optimizing multi-qubit gates in the setting of circuit quantum electrodynamics. We will consider both the standard two-qubit setup, as well as the three-qubit configuration where interactions exist only between qubits 1 and 2 and qubits 2 and 3, while the gate is to be applied between qubits 1 and 3. Note that the latter situation is of far more general relevance than the cavity grid only. In fact, it will become important whenever full global coupling (all qubits couple to all others) is not available, which is the expected situation for multi-qubit circuits. Optimal control techniques have already been successfully applied to other three-qubit architectures, in both analytical \cite{KGB3Qubit, HaidongSpinChain, KhanejaSpoerl3Qubit}, and numerical\cite{Neves3Qubit} approaches.

The remainder of the paper is organized as follows: We first provide more background on the cavity grid architecture and our optimal control methods in Section~\ref{methods}. Afterwards, we go through a sequence of models of increasing sophistication, for which we discuss the results found using optimal control. We look at two qubits in one cavity and three qubits in two cavities (the cavity grid situation). At first, in Section~\ref{idealized}, we consider both of these situations in the setting of an {}`idealized model' where we assume local control over all the qubits. Later on in Section~\ref{realistic}, we introduce a {}`realistic model', in which we take into account that a microwave pulse applied through a resonator will couple to both qubits, such that some amount of local control is lost. In addition, in the {}`realistic model', we will consider restrictions on the pulse amplitudes. In all cases, we are interested in finding the minimum time at which the optimized pulse sequence has fidelity unity, and we extract the speedup vs. the known, sequential approach. We will discuss typical control sequences and also present the evolution of entanglement between the qubits during an optimized three-qubit scheme.
\section{Methods \& Setup}
\label{methods}
\subsection{The cavity grid}
We briefly recount the basic setup of the two-dimensional cavity grid\cite{CavityGridEPL}. The idea is to have a grid of transmission line resonators (horizontal and vertical, in two different layers), and to place qubits at the intersections, as depicted in Fig.~\ref{cavGrid}. Each qubit thus feels the microwave field of the two cavities crossing at its location. It thus couples directly, via the standard dispersive $XY$-type interaction $\sigma_{1}^{x}\sigma_{2}^{x}+\sigma_{1}^{y}\sigma_{2}^{y}$ (see Section~\ref{idealized}), to all of the qubits in both of these cavities. Qubit frequencies are chosen distinct to provide for individual addressability. i{\sc swap} gates can then be implemented by bringing two qubits into mutual resonance (still detuned from the coupling cavity), evolving for an appropriate time, and bringing them out of resonance again. For an $n\times n$ array of $n^{2}$ qubits, only $n$ different frequencies are needed, which drastically eases the restrictions for the size of the array vs. the one-dimensional case.

A crucial question, however, is how to couple two qubits that are not part of the same cavity. In nearest-neighbor coupled arrays, this would require a number of intermediate {\sc swap} operations that grows with the size of the array. In contrast, the overhead remains constant for the case of the cavity grid. Assuming direct couplings exist between $1-2$ and $2-3$, then any two-qubit gate between $1$ and $3$ can be implemented by first swapping the quantum information from $1$ to $2$, then performing the desired operation, and finally swapping back. Thus, since any two qubits will share a third qubit at the corner
of the cavities to which they couple, this provides a general, distance-independent coupling scheme.

However, one of the disadvantages of such a straightforward pulse sequence is that the {\sc swap} gate itself is not elementary with respect to the $XY$ interaction, which only yields an i{\sc swap}. Unfortunately, three i{\sc swap}s are needed to compose one {\sc swap}, of which two are necessary
for the {}`coupling around the corner' approach. Thus, this very important operation in the cavity grid architecture lends itself naturally to potentially significant improvements via optimal control techniques. 
\subsection{Our optimal control approach}
\label{optContOverview}
The general framework in optimal control is to maximise a figure of merit subject to steering a dynamic system according to its equation of motion under experimentally admissible controls. In quantum information processing, a convenient figure of merit often chosen is the trace fidelity of the gate $U(T)$ actually synthesised under the available controls at some final time $T$ with respect to the desired target unitary gate $U_{\rm target}$. For $n$ qubits and setting $N:=2^n$ the (squared) trace fidelity is
\begin{eqnarray*}
F^2_{\rm tr}\; :=\;  | \tfrac{1}{N}\;  \tr \{U_{\rm target}^\dagger U(T)\}\;|^2\,.
\end{eqnarray*}
Now, for closed quantum systems Schr{\"o}dinger's equation of motion lifted to operator form reads
\begin{eqnarray*}
\dot U \; = \; -i H U\,,
\end{eqnarray*}
where the total Hamiltonian $H$ is composed of the non-switchable drift term $H_d$ 
and control terms $H_j$ governed by piecewise constant control amplitudes $u_j(t_k)$
for $t_k\in[\,0,T]$ so as to give
\begin{eqnarray*}
H(t_k) := H_d + \sum_j u_j(t_k) H_j \,.
\end{eqnarray*}
Dividing the total time $T$ into $M$ intervals $\Delta t_k$ (of piecewise constant controls) with $T=\sum_k^M \Delta t_k$, the gate $U(T)=U_M\, U_{M-1}\cdots U_k\cdots U_2\, U_1$ is made of components
$U_k = \exp\{-i H(t_k)\Delta t_k\}$. Then the derivatives
\begin{eqnarray*}
\begin{split}
\tfrac{\partial F^2_{\rm tr}}{\partial u_j(t_k)}=& \tfrac{-2}{N^2}\Re\Big(\tr\!
\big\{ U^\dagger_{\rm target} \, U_M \, U_{M-1} \cdots U_{k+1}\times\nonumber\\
& \big(i \Delta t_k H_j U_k \big) U_{k-1} \cdots U_2 \, U_1 \big\} \tr\{U^\dagger_{\rm target} U(T)\}^*\Big)
\end{split}
\end{eqnarray*}
can (with appropriate step size $\alpha_r >0$) readliy be used for recursive gradient schemes like
\begin{eqnarray}
\label{gradUpdate}
u_j^{(r+1)}(t_k) = u_j^{(r)}(t_k) + \alpha_r \tfrac{\partial F^2_{\rm tr}}{\partial u_j(t_k)}
\end{eqnarray}
representing the simple setting of steepest ascent. Likewise, conjugate gradient or Newton methods can be
implemented \cite{GRAPE,LBFGS}. Here we used the {\sc lbfgs} variant of a quasi-Newton approach, as sketched in Appendix \ref{numerics}. We refer to algorithms of this kind as Gradient Ascent Pulse Engineering\cite{GRAPE} ({\sc grape}) algorithms.

Spin and pseudo-spin systems are a particularly powerful paradigm of quantum systems, in particular when they are fully operator controllable, i.e.~universal. For this to be the case the drift and the control Hamiltonians have to generate the full $n$-qubit unitary algebra $\mathfrak{su}(2^n)$ by way of commutation \cite{SJ72JS,TOSH-Diss,AlbAll02}. A simple example exploited in the {\sc stirap}\cite{STIRAP} scheme is the Lie algebra of local control $\mathfrak{su}(2)$ being generated by the Pauli matrices $\sigma_x$ (pulse) and $\sigma_z$ (detuning), whose commutator gives $\sigma_y$ and thus introduces phase sensitivity. In the instance of the (realistic) two-qubit Hamiltonian examined in Section~\ref{realistic} (Eqn.~\ref{twoQubitRealisticHam}), only different phase shifts $\Delta_1 \neq \Delta_2$ ensure full controllability. In practice for universality it suffices that ({\em i\,}) all qubits can be addressed selectively and ({\em ii\,}) that they form an arbitrary connected graph of Ising-type coupling interactions. More recent analyses revealed that even partially collective controls maintain universality in different types of Ising or Heisenberg coupled systems as long as the hardware architecture gives rise to Hamiltonians with no symmetries\cite{SS09}.

\section{Idealized Model}
\label{idealized}
In order to establish the lower limits on gate times in this scheme, let us first consider an idealized model where the control fields are unrestricted. A model which respects more closely the limitations in current experiments will be considered in the following section. After adiabatic elimination of the cavity mode\cite{Blais2004}, the effective qubit-qubit interaction Hamiltonian is of the form
\begin{eqnarray*}
H_{\text{int}} &=& \pi J\left(\sigma^{+}_{1}\sigma^{-}_{2} +\sigma^{-}_{1}\sigma^{+}_{2}\right)\nonumber\\
&=&\frac{\pi J}{2}\left(\sigma^{x}_{1}\sigma^{x}_{2}+\sigma^{y}_{1}\sigma^{y}_{2}\right),
\end{eqnarray*}
where $J$ is an effective coupling constant determined by the qubit-cavity couplings and detunings. Evolution under $H_{\text{int}}$ for a time of $T=\frac{1}{2J}$ yields the so-called i{\sc swap} operation\cite{SchuchSiewert}:
\begin{eqnarray}
\label{iSwap}
\exp\left\{-i\frac{1}{2J}H_{\text{int}}\right\}=
\left(\begin{array}{cccc} 1 & 0 & 0 & 0 \\ 0 & 0 & -i & 0 \\ 0 & -i & 0 & 0 \\ 0 & 0 & 0 & 1 \end{array}\right),
\end{eqnarray}
a universal two-qubit gate which can be considered the `natural' gate of the coupling interaction.
\subsection{Two qubits in a cavity}
\label{twoQubitIdeal}
If the two coupled qubits are individually addressable by resonant microwave fields of tunable amplitude and phase, the total Hamiltonian in a frame rotating with the driving fields is
\begin{eqnarray*}
H_{\text{ideal}}^{(2)}(t)&=&H_{\text{int}}
+\sum_{i=1}^{2}\pi\left(\,\Omega^{x}_i(t)\,\sigma^x_{i}+\Omega^{y}_i(t)\,\sigma^y_{i}\,\right),
\end{eqnarray*}
under the assumption that the two qubits are in resonance (i.e. they are set to the same frequency, but distinct from the cavity frequency). Note that our simulations work within the rotating-wave approximation
and assume a sufficiently detuned cavity that has already been eliminated. Thus, we neglect both the Bloch-Siegert shift that would arise for extremely strong driving, as well as any AC Stark shift due to a strong cavity population. In order to make the bilinear control form of the Hamiltonian more explicit, the microwave fields are specified in terms of real and imaginary parts $\Omega^{x}_i(t)$ and $\Omega^{y}_i(t)$, respectively, rather than amplitude and phase. Throughout this article the controls and $J$-coupling are normalized as frequencies rather than angular frequencies, with the $2\pi$ factors written explicitly in the Hamiltonians (and with $\hbar=1$).

\begin{figure}[!htbp]
\includegraphics[width=0.97\columnwidth]{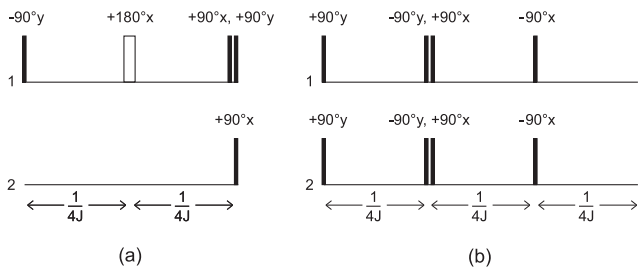}
\caption{Analytical pulse sequences for time-optimal implementations of two-qubit gates: (a) the {\sc cnot} gate, where 1 is the control qubit and 2 is the target qubit, and (b) the {\sc swap} gate.}
\label{analyticalSequences}
\end{figure}

One approach to implement a general two-qubit gate is to decompose it into a sequence of i{\sc swap} gates and local operations, as discussed in Refs.~\onlinecite{SchuchSiewert,CavityGridEPL}. For example, a {\sc cnot} gate can be created from two i{\sc swap}s, while a {\sc swap} gate requires three. We refer to this as the `sequential' approach. In this section we assume that local operations can be performed in a negligible time compared to the time required by the coupling evolution. Time-optimal pulse sequences for an arbitrary two-qubit gate can then be determined analytically via the Cartan decomposition of $\text{SU}(4)$ as in Refs.~\onlinecite{Cartan, CartanS2S}. The i{\sc swap} implementation suggested in Eqn.~(\ref{iSwap}) is, unsurprisingly, already time-optimal. Time-optimal pulse sequences for the {\sc swap} and {\sc cnot} are provided in Fig.~\ref{analyticalSequences}. A comparison of the times required by the different schemes is given in Table~\ref{table1} - we find that even in this simple case the {\sc swap} and {\sc cnot} can be sped up by a factor of 2.
\subsection{Three qubits in two cavities}
\begin{figure}[!htbp]
\includegraphics[width=0.97\columnwidth]{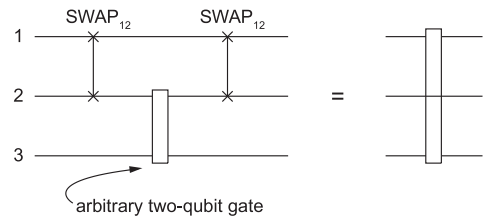}
\caption{The standard decomposition of an indirect two-qubit gate into direct two-qubit gates via the {\sc swap} operation.}
\label{swapPrinciple}
\end{figure}

We now consider two qubits, each in a separate cavity, which are coupled indirectly via an additional `mediator' qubit placed at the intersection of the cavities. If local controls on all three qubits are available, the Hamiltonian is
\begin{eqnarray*}
H_{\text{ideal}}^{(3)}(t)&=&\frac{\pi J}{2}\left(\sigma^{x}_{1}\sigma^{x}_{2}+\sigma^{y}_{1}\sigma^{y}_{2}
+\sigma^{x}_{2}\sigma^{x}_{3}+\sigma^{y}_{2}\sigma^{y}_{3}\right)\nonumber\\
&+&\sum_{i=1}^{3}\pi\left(\Omega^{x}_i(t)\,\sigma^x_{i}+\Omega^{y}_i(t)\,\sigma^y_{i}\right).\\
\end{eqnarray*}
Gates can be implemented between the indirectly coupled qubits (1 and 3) in the sequential scheme via the {\sc swap} operation, as depicted in Fig.~\ref{swapPrinciple}. For example, an i{\sc swap} between qubits 1 and 3 could be implemented as a sequence of seven two-qubit i{\sc swap}s.

\begin{figure}[!hbtp]
\includegraphics[width=0.97\columnwidth]{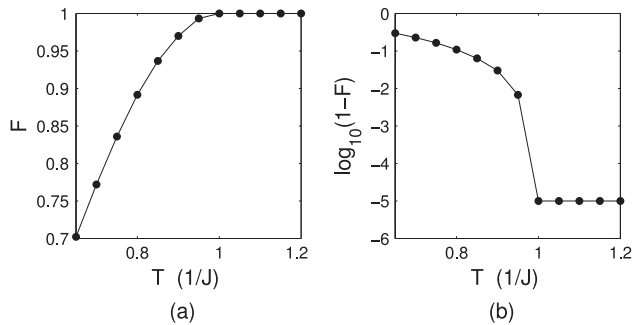}
\caption{(a) Maximum achievable fidelity as a function of pulse duration in the three-qubit idealized model for an i{\sc swap}$_{13}$ gate. (b) On a logarithmic scale we observe a sharp convergence to the threshold fidelity of $1-10^{-5}$, where the algorithm terminates.}
\label{iSwapIdealTop}
\end{figure}

However, these indirect two-qubit gates embedded in a three-qubit system can be implemented considerably faster using optimized controls. The analytical methods used for determining time-optimal two-qubit gates cannot be applied here; instead we use the numerical techniques outlined in Section~\ref{methods}. For a fixed gate time, an initial pulse is chosen and iterated until the {\sc grape} algorithm converges to a maximum of the fidelity. This procedure is repeated for a range of different gate times, allowing us to estimate the minimal time. Further details about the numerics are included in Appendix \ref{numerics}. A plot of maximum fidelity vs. gate time for the case of an i{\sc swap}$_{13}$ gate is shown in Fig.~\ref{iSwapIdealTop}. In this case we find that a time of $1/J$ is required to reach the threshold fidelity. Minimal times for other indirect two-qubit gates are similarly calculated and the results are included in Table~\ref{table1}, alongside the times required by the corresponding sequential schemes of decomposition into two-qubit i{\sc swap}s.
\begin{table}
\centering
\begin{tabular}{c c c c}
\hline
$\:$ Gate $\:$ & $\:$ $T_{\text{seq}}$ ($1/J$) $\:$ & $\:$ $T_{\text{opt}}$ ($1/J$) $\:$
& $\:$ speedup factor $\:$ \\
\hline \hline
$\:$ i{\sc swap}$_{12}$ $\:$ & $\:$ $0.5$ $\:$ & $\:$ $0.5$ $\:$    & $\:$ -   $\:$	 \\
$\:$ {\sc cnot}$_{12}$  $\:$ & $\:$ $1.0$ $\:$ & $\:$ $0.5$ $\:$    & $\:$ $2$ $\:$	 \\
$\:$ {\sc swap}$_{12}$  $\:$ & $\:$ $1.5$ $\:$ & $\:$ $0.75$ $\:$   & $\:$ $2$ $\:$	 \\
\hline
$\:$ i{\sc swap}$_{13}$ $\:$ & $\:$ $3.5$ $\:$ & $\:$ $1.00^*$ $\:$ & $\:$ $3.50$ $\:$ \\
$\:$ {\sc cnot}$_{13}$  $\:$ & $\:$ $2.0$ $\:$ & $\:$ $1.00^*$ $\:$ & $\:$ $2.00$ $\:$ \\
$\:$ {\sc swap}$_{13}$  $\:$ & $\:$ $4.5$ $\:$ & $\:$ $1.15^*$ $\:$ & $\:$ $3.91$ $\:$ \\
\hline
\end{tabular}
\caption{Implementation times for a selection of direct and indirect two-qubit gates in the \textbf{idealized} model: $T_{\text{seq}}$ is the time required by decomposing the gate into two-qubit i{\sc swap}s; $T_{\text{opt}}$ is the time required by the optimal control sequence. The times marked with an asterisk are determined numerically as the shortest times in which the {\sc grape} algorithm can reach a fidelity of $1-10^{-5}$, with time resolution $0.05/J$. The time of $2.0/J$ for the sequential implementation of a {\sc cnot}$_{13}$ is a special case, where the two {\sc swap}s in Fig.~\ref{swapPrinciple} can be replaced by i{\sc swap}s\cite{CavityGridEPL}.}
\label{table1}
\end{table}

\begin{figure}[!htbp]
\includegraphics[width=0.97\columnwidth]{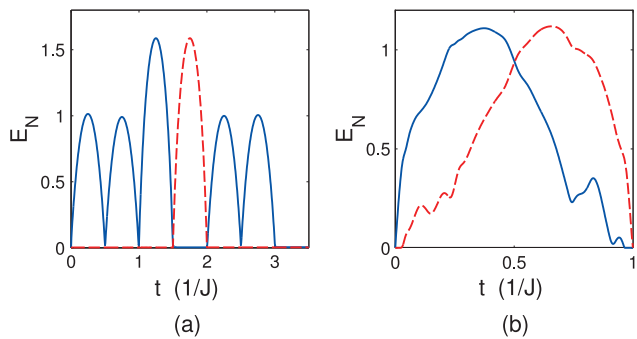}
\caption{(Color online) Logarithmic negativity between qubit pairs 1-2 (blue line) and 2-3 (red, dashed line) for (a) sequential and (b) optimized implementations of an i{\sc swap}$_{13}$.}
\label{entPlots}
\end{figure}

In order to illustrate how the optimized indirect two-qubit gates differ from the sequential schemes, we can consider the entanglement between directly coupled qubits over the time interval during which the controls are applied. For this we use the logarithmic negativity\cite{plenioLogNeg}, defined as
\begin{eqnarray*}
E_{N}(\rho)=\log_{2}\left|\left|\rho^{\Gamma_{A}}\right|\right|_{1}
\end{eqnarray*}
where $\Gamma_{A}$ is the partial transpose, $\left|\left|\cdot\right|\right|_{1}$ is the trace norm, and $\rho$ is the reduced density matrix of the two-qubit subsystem. We choose the initial state $|100\rangle$ and apply the i{\sc swap}$_{13}$ operation while observing the entanglement between qubit pairs 1-2 and 2-3, illustrated in Fig.~\ref{entPlots}. In the sequential scheme the mediator qubit is entangled either with qubit 1 or qubit 3. In the optimized case, as one might expect, the mediator qubit is simultaneously entangled with both.
\section{Realistic Model}
\label{realistic}
Allowing for unrestricted $x$ and $y$ control on each qubit yields lower bounds on what implementation times are possible. However we would also like to consider a restricted model of coupled superconducting qubits which is more feasible in current experiments. We allow for individual tuning of the qubit resonance frequencies ($z$ controls), but restrict ourselves to a single microwave field ($x$ control) per cavity, where the microwave field is no longer required to have tunable phase.
\subsection{Two qubits with restricted controls}
\begin{figure}[!htbp]
\includegraphics[width=0.97\columnwidth]{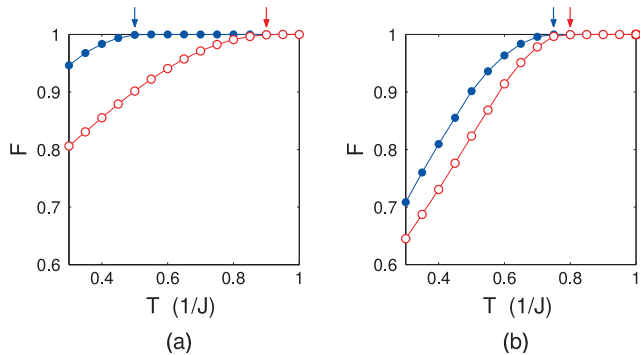}
\caption{(Color online) Maximum fidelity as a function of pulse duration in the two-qubit realistic model for (a) a {\sc cnot} gate, and (b) a {\sc swap} gate. Maxima obtained with no restrictions on the controls are shown in blue ($\bullet$), while those obtained under the restrictions in Eqn.~(\ref{contRestrictions}) are shown in red ($\circ$). The arrows indicate the minimal times at which the threshold fidelity is achieved.}
\label{twoQubitRealisticTop}
\end{figure}

\begin{figure}[!htbp]
\includegraphics[width=0.97\columnwidth]{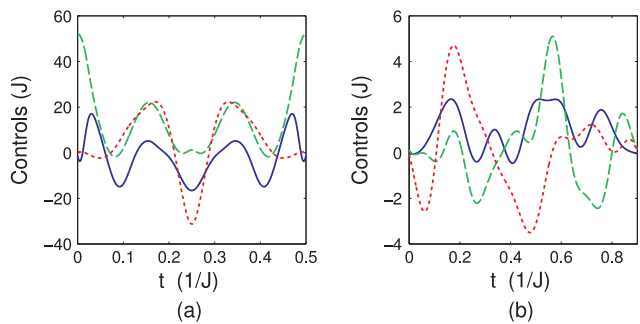}
\caption{(Color online) Sample controls obtained by the {\sc grape} algorithm for the minimal-time implementation of a {\sc cnot} in the two-qubit realistic model: $\Delta_{1}$ (red, dotted line), $\Delta_{2}$ (green, dashed line), and $\Omega$ (blue, solid line) with (a) unrestricted controls, and (b) restricted controls. Note that the time course of the controls in (a) is perfectly palindromic as the controls may
equally well be read forward and backward in time.}
\label{twoQubitRealisticControls}
\end{figure}

Under these structural restrictions, the corresponding two-qubit Hamiltonian is
\begin{eqnarray}
\label{twoQubitRealisticHam}
H_{\text{real}}^{(2)}(t)&=&H_{\text{int}}\!+\!\sum_{i=1}^{2}\pi\Delta_{i}(t)\,\sigma^z_{i}
+\pi\Omega(t)\left(\sigma^{x}_{1}+\sigma^{x}_{2}\right)\quad
\end{eqnarray}
where $\Omega(t)$ is the amplitude of the microwave field and $\Delta_{i}(t)$ are the detunings of the qubit frequencies from the microwave carrier frequency. We consider two possible cases for the control functions: ({\em i\,}) the controls are unrestricted, or ({\em ii\,}) the controls are restricted to the following ranges:
\begin{eqnarray}
\label{contRestrictions}
|\Delta_{i}(t)| &\leq& \Delta_{\text{max}} =1000\text{MHz}\nonumber\\
|\Omega(t)|     &\leq& \Omega_{\text{max}} =50\text{MHz},
\end{eqnarray}
where the coupling constant was taken to be $J=21\text{MHz}$ (as in Ref.~\onlinecite{CavityGridEPL}) in our numerical examples. Furthermore, in case ({\em ii\,}) we require that the controls start and end at zero with a maximum rise-time of $4\rm{ns}$, which should be feasible in current experiments~\cite{Yale2Qubit2009}. In case ({\em i\,}) the results from Section~\ref{twoQubitIdeal} still apply - we need only to rewrite the local $x$ and $y$ pulses in terms of our new controls. For instance a $90^{\circ}$ $x$-rotation on the first qubit can be decomposed as
\begin{eqnarray*}
R_{1}^{x}\!\left(90^{\circ\!}\right)=R_{2}^{z}\!\left(-180^{\circ\!}\right)
R_{1,2}^{x}\!\left(45^{\circ\!}\right)R_{2}^{z}\!\left(180^{\circ\!}\right)
R_{1,2}^{x}\!\left(45^{\circ\!}\right)
\end{eqnarray*}
and the other local $x$ and $y$ pulses can be similarly decomposed. Thus, the two-qubit times in Table~\ref{table1} also hold for this case. In particular, this means that case ({\em i\,}) also yields the optimum times for an experimental setup with both $x$ and $y$ pulses available (variable phase of the microwave drive) and without restrictions on the local gate controls. In case ({\em ii\,}) the analytical methods are no longer applicable, as they require that local rotations can be applied in negligible time. The i{\sc swap} can of course still be implemented by simply evolving under the coupling, but to find time-optimal implementations for other two-qubit gates we again apply the {\sc grape} algorithm. This can be forced to optimize only over controls in a certain amplitude or bandwidth range~\cite{GRAPE,PRL_decoh2}. Fig.~\ref{twoQubitRealisticTop} contains the fidelity vs. pulse duration curves for two-qubit {\sc swap} and {\sc cnot} gates. Examples of the optimized controls obtained by the {\sc grape} algorithm for the {\sc cnot} gate are provided in Fig.~\ref{twoQubitRealisticControls}.

Observe that the {\sc cnot} gate is self-inverse and the two-qubit Hamiltonian in Eqn.~(\ref{twoQubitRealisticHam}) is real and symmetric. As we described earlier \cite{Spoerl2007}, in such systems there may be {\em palindromic} control sequences as in Fig.~\ref{twoQubitRealisticControls}a. Their practical advantage lies in the fact that they may be synthesized on $LCR$ terminals only using capacitances ($C$) and inductances ($L$) and no resistive elements ($R$) thus avoiding losses. In contrast, since the i{\sc swap} is only a fourth root of the identity, it is no longer self-inverse, and therefore palindromic controls are not to be expected (compare, e.g., Fig.~\ref{threeQubitControls}.)
\subsection{Three qubits with restricted controls}
\begin{figure}[!htbp]
\includegraphics[width=0.97\columnwidth]{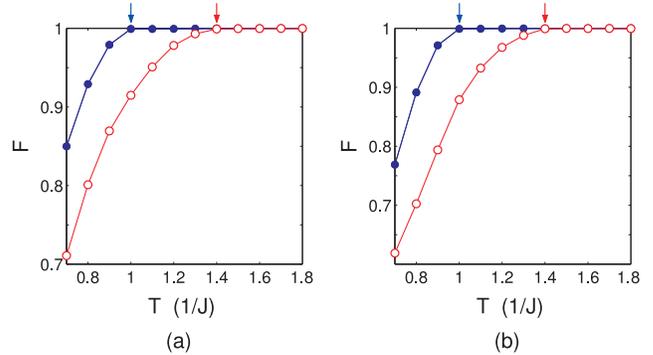}
\caption{(Color online) Maximum fidelity as a function of pulse duration in the three-qubit realistic model for (a) a {\sc cnot}$_{13}$ gate, and (b) an i{\sc swap}$_{13}$ gate. Maxima obtained with no restrictions on the controls are shown in blue ($\bullet$), while those obtained under the restrictions in Eqn.~(\ref{contRestrictions}) are shown in red ($\circ$). The arrows indicate the minimal times at which the threshold fidelity is achieved.}
\label{threeQubitTop}
\end{figure}

\begin{figure}[!htbp]
\includegraphics[width=0.97\columnwidth]{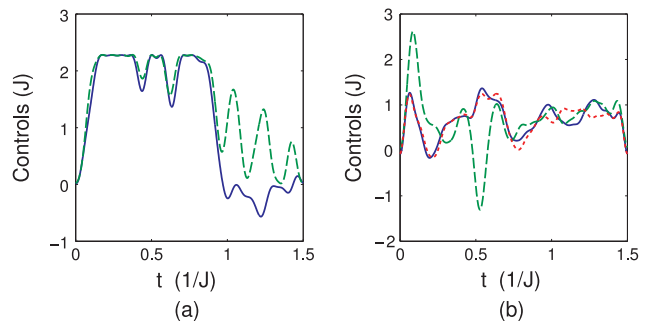}
\caption{(Color online) Sample controls to implement an i{\sc swap}$_{13}$ gate in the three-qubit realistic model with restrictions ({\em ii\,}) in place: (a) $\Omega^{x}_{12}$ (green, dashed line), $\Omega^{x}_{23}$ (blue, solid line). (b) $\Omega^{z}_{1}$ (blue, solid line), $\Omega^{z}_{2}$ (green, dashed line), $\Omega^{z}_{3}$ (red, dotted line).}
\label{threeQubitControls}
\end{figure}

For three qubits coupled via two cavities we allow for three local $z$ controls and two $x$ controls, with the Hamiltonian
\begin{eqnarray*}
H_{\text{real}}^{(3)}(t)&=&\frac{\pi J}{2}\left(\sigma^{x}_{1}\sigma^{x}_{2}+\sigma^{y}_{1}\sigma^{y}_{2}
+\sigma^{x}_{2}\sigma^{x}_{3}+\sigma^{y}_{2}\sigma^{y}_{3}\right)\nonumber\\
&+&\pi\Omega^{x}_{12}(t)\left(\sigma^x_{1}+\sigma^x_{2}\right)
+\pi\Omega^{x}_{23}(t)\left(\sigma^x_{2}+\sigma^x_{3}\right)\nonumber\\
&+&\sum_{i=1}^{3}\pi\Omega^{z}_i(t)\,\sigma^z_{i}.
\end{eqnarray*}
Again we determine optimized controls numerically; fidelity vs. pulse duration curves are shown in Fig.~\ref{threeQubitTop}, while sample optimized controls for the restricted case ({\em ii\,}) are shown in Fig.~\ref{threeQubitControls}. Only the $x$ restriction plays a role here as the $z$ restriction is an order of magnitude larger. A comparison of times in the sequential and optimized schemes under the restrictions in Eqn.~(\ref{contRestrictions}) is provided in Table~\ref{table2}. The times in the sequential scheme have increased, as each local $90^{\circ}$ $x$-rotation now requires a time of $0.25/\Omega_{\text{max}}=0.105/J$. The times required by the optimized schemes also increase, but substantial speedups are still possible.
\begin{table}
\centering
\begin{tabular}{c c c c}
\hline
$\:$ Gate $\:$ & $\:$ $T_{\text{seq}}$ ($1/J$) $\:$ & $\:$ $T_{\text{opt}}$ ($1/J$) $\:$
& $\:$ speedup factor $\:$ \\
\hline \hline
$\:$ i{\sc swap}$_{12}$ $\:$ & $\:$ $0.50$  $\:$ & $\:$ $0.50$ $\:$	& $\:$ - $\:$\\
$\:$ {\sc cnot}$_{12}$  $\:$ & $\:$ $1.21$  $\:$ & $\:$ $0.90$ $\:$	& $\:$ $1.34$ $\:$\\
$\:$ {\sc swap}$_{12}$  $\:$ & $\:$ $1.82$ $\:$ & $\:$ $0.80$ $\:$	& $\:$ $2.28$ $\:$\\
\hline
$\:$ i{\sc swap}$_{13}$ $\:$ & $\:$ $4.13$  $\:$ & $\:$ $1.40$ $\:$	& $\:$ $2.95$ $\:$\\
$\:$ {\sc cnot}$_{13}$  $\:$ & $\:$ $2.21$  $\:$ & $\:$ $1.40$ $\:$	& $\:$ $1.58$ $\:$\\
\hline
\end{tabular}
\caption{Implementation times for a selection of direct and indirect two-qubit gates in the \textbf{realistic} model with the control amplitudes restricted as described in case ({\em ii\,}). $T_{\text{seq}}$ is the time required by decomposing the gate into two-qubit i{\sc swap}s and local operations; $T_{\text{opt}}$ is the time required by the numerically optimized pulse to reach a fidelity of $1-10^{-3}$. The particular values for the minimal times simply result from our choice for the maximum amplitude relative to $J$.}
\label{table2}
\end{table}
\section{Conclusions} 
We have demonstrated how optimal control methods provide fast high-fidelity quantum gates for coupled superconducting qubits. In contrast to conventional approaches that make use of the coupling evolutions sequentially (i.e., along one dimension at a time), numerical optimal control exploits the coupling dimensions simultaneously thereby gaining significant speedups. In particular, our numerical method provides constructive controls under realistic experimental conditions, such as ({\em i\,}) power and rise-time limits in the control amplitudes, ({\em ii\,}) additional individual detunings on each qubit, and ({\em iii\,}) lack of phase switching and restriction to controls affecting qubits jointly. We showed how the latter two complement one another. Finally, our approach to providing optimized controls tailored to the hardware of 2D cavity grids scales to (2D) arrays of many qubits and arbitrary gate operations between two generic qubits picked from such grids. It is therefore anticipated to find wide application in similar architectures of large-scale quantum systems used as quantum simulators, processors, or storage devices.
\begin{acknowledgments}
This work was supported in part by the {\sc eu} programs {\sc qap} and Euro{\sc sqip}, by the Bavarian network {\sc enb} via the international PhD program {\em Quantum Computing, Control, and Communication} ({\sc qccc}), by the Deutsche Forschungsgemeinschaft ({\sc dfg}) in the collaborative research centre {\sc sfb} 631, via the Nanosystems Initiative Munich ({\sc nim}), and via the Emmy-Noether program. We wish to thank Ilya Kuprov and Uwe Sander for making available a fast {\sc lbfgs}-version of {\sc grape} ({\em see below}).
\end{acknowledgments}
\appendix
\section{Numerical details}
\label{numerics}
The gradients introduced in Section~\ref{optContOverview} form the basis of our numerical optimization algorithm, but there is still some freedom in how they can be used to update the controls. The simplest approach of adding the gradients directly to the controls with some positive stepsize, as in Eqn.~(\ref{gradUpdate}), is typically not the most numerically efficient one. Here we replace the conjugate-gradient updates of our standard routine\cite{GRAPE} by a Newton method which makes use of the second derivative of the quality function (the Hessian matrix). Because we chose to divide our evolution into $M=256$ constant intervals, this results in a $256\times 256$ Hessian matrix, direct computation of which would be inefficient. Instead we use the standard limited-memory Broyden-Fletcher-Goldfarb-Shanno ({\sc lbfgs}) algorithm to approximate the Hessian matrix by analyzing successive gradient vectors\cite{LBFGS,fmincon}.

Inherent to these gradient-based methods is that they result only in a local maximum of the fidelity. To increase confidence that optima close to the global maximum will be obtained, we can sample from a range of random initial controls. For the results presented in this paper we used the following procedure:
\begin{enumerate}
\item[({\em i\,})] Randomly generate 50 initial controls and optimize each in 100 iterations of the {\sc grape} algorithm.
\item[({\em ii\,})] Select the highest 10 fidelities and iterate each a further 500 times.
\item[({\em iii\,})] Select the highest 2 fidelities and iterate each a further 1000 times.
\end{enumerate}
This procedure is applied for a range of pulse durations, yielding an estimate of the minimum time required to implement the given operation.
\bibliography{RobFlorianRefs,ThomasRefs}
\end{document}